# Institutional Policy Pathways for Supporting Research Software: Global Trends and Local Practices

Authors: Michelle Barker, Jeremy Cohen, Pedro Hernández Serrano, Daniel S. Katz, Kim Martin, Dan Rudmann, Hugh Shanahan

## Abstract

Research software is essential to modern science, yet many research-performing organisations lack coherent policies to support its development, sustainability, and recognition. Despite its central role in research outcomes, research software and its personnel are often excluded from research institution policies. This article discusses the work of the Policies in Research Organisations for Research Software (PRO4RS) Working Group, exploring current gaps, including limited support for research software personnel, and offering recommendations for embedding software into policy frameworks to ensure the software is valued, sustained, and aligned with broader research goals. The analysis proposes a three-layer framework to guide policy development: central policies that explicitly recognise software as a scholarly output; middle-layer policies that align related areas such as open science, intellectual property, and research evaluation; and outer-layer mechanisms like guidelines and frameworks that enable practical implementation. Institutions are encouraged to assess existing practices, adopt international declarations, and engage stakeholders to advance software recognition. Stronger institutional policies can foster good practices, boost collaboration, support reproducibility, and strengthen researcher development, to maximise both institutional value and research impact, and position organisations as leaders in open, sustainable, software-driven science.

## 1. Introduction

Research software is foundational to contemporary science and it is of critical importance that it is built to be robust, sustainable and maintainable, using widely accepted best practices. It is therefore vital that research performing organisations (RPOs) worldwide develop, align, and implement policies on research software, as a key component of open science. Excellence in this area empowers institutions to deliver stronger research impact than those that take a different path. Institutions often lack coherent strategies to manage research software assets, provide clear guidance, or embed recognition of software development into hiring, promotion, and research evaluation frameworks. This lack of institutional focus has implications for sustainability, legal clarity, data integrity, and reproducibility. Furthermore, it affects the visibility and career development of research software personnel, who are critical contributors to the research ecosystem. However, institutional policies that govern, support, and recognise research software and its creators remain underdeveloped across many RPOs.

Helping to analyse and address this, and to develop examples and recommendations for institutional policies has been a primary aim of the Policies in Research Organisations for Research Software (PRO4RS) Working Group (WG). This article broadly explores the context and implications of this work to make recommendations on approaches that can be taken to support policy change. In scientific research, data management policies were introduced to promote transparency, uphold integrity, ensure preservation, and maximise the value of research data, but the introduction of the similarly required research software policies has been much more limited. While many research institutions include research software in policies on infrastructure, licensing, or open science, few include it in areas concerning research assessment reform, an area that is critical for recognising and rewarding research software as a primary research output.

Yet there are a range of broader initiatives around reform of research assessment that are championing recognition of a range of open science outputs, including software and code; alongside practical implementation by a small number of research institutions. This article concludes with recommendations on how to utilise policy to improve recognition of the important role software now plays in research and to ensure that the investment in people, skills and infrastructure to support research software development results in sustainable, maintainable software assets that produce reproducible outputs.

## 2. The Need for Policies in Research Organisations That Support Research Software

As research across almost all domains increasingly relies on research software, there is a need for better recognition of the importance and value of that software to modern research outputs (1). An increasing number of initiatives are focusing on implementation of research software within the context of FAIR (Findable, Accessible Interoperable, Reusable) and open science. Efforts to align policies from funders, journals, and governments are also in progress. However, there is a significant gap around how RPOs can develop and implement local policies to help support sustainability of software outputs, ensure reproducibility of research outputs, and underpin research quality.

To help develop this improved recognition, and to manage a number of other aspects related to the use of software to support and undertake research, there is a strong need for institutions to provide guidance and policies relating to research software. This can assist in achieving many institutional aims, including to promote good research practices, enhance collaboration, foster reproducibility, and maximise the value and impact of efficient and reliable scientific research. Excellence in this area can also enable institutions to achieve greater research impact than peer organisations who do not have this approach. Software policies can cover multiple aspects of software management; namely, its development, legal, ethical and secure use, protection of intellectual property rights, interaction with user and developer communities and how research software engineering as a skill is used in hiring, evaluation and promotion

The need for research institutions to better manage (and leverage) their research software assets can be demonstrated by the increasing data on how much research software is developed by research institutions, with examples including:

- The number of public GitHub repositories that University of California San Diego faculty and students have contributed to was estimated to be 32,000 (2).
- A Survey of Open Source Software Repositories in the US Department of Energy's National Laboratories identified 2,005 repositories, of which just over half were updated within the previous 12 weeks (3).
- The French Research Software Catalogue lists software developed in French research labs (4), and included more than 1300 software references as of mid-2025 (5).
- Software Heritage's OSPO-RADAR project maps and monitors the landscape of Open Source Program Offices (OSPOs) at research institutions to raise the profile of research software (6).
- Nearly half of Australian Research Council (ARC) grants from 2010-2019 resulted in software production, highlighting the need to strengthen research software support (7).

One of the main challenges to the widespread development and adoption of research software policies at research organisations is a lack of existing policy-related activity in relation to research software. The need to better support and recognise research software and its personnel is increasingly recognised at national levels (8–10) and has been well articulated (8–10) in a range of national policies. However, institutional focus on how to effectively manage and sustain research software outputs is also essential. Many institutions now provide some sort of guidance in areas such as open source licensing, but much more work is needed to ensure policies address the breadth of areas where research software and its personnel should be recognised and supported.

The inclusion of software and source code in definitions of open science has also resulted in its inclusion in many institutional policies on supporting open science; however, these usually mention software only in passing, and then do not address how software requires different types of recognition and support to other elements of research. In contrast, research data management policies have been widely implemented to encourage transparency, integrity, preservation, and to maximise the value of research data. Whilst research software policies are closely interconnected to and mutually supportive of research data management policies, research software is fundamentally different from research data and needs to be treated as such (11). Research software (and other research outputs) are not simply another type of data.

### 3. The Policies in Research Organisations for Research Software (PRO4RS) WG

The PRO4RS WG was a joint initiative of the Research Software Alliance (ReSA) and the Research Data Alliance (RDA) from 2023 to 2025. The PRO4RS WG aimed to advance the ability of RPOs worldwide to create, share, manage, and re-use research software (12). The WG undertook work designed to support RPOs in developing, aligning, and implementing policies on research software, as a key component of FAIR and open science. The WG, led by eight co-chairs, undertook activities involving more than 100 participants (including PRO4RS WG members) and that included engagement through ReSA and RDA events and channels over an 18-month period.

The PRO4RS WG produced a range of outputs that research organisations can utilise to improve support for research software and its personnel through organisational policy:

1. A list of publicly-available institutional policies that support research software, from around the globe.
2. A report summarising resources relevant to different stakeholders on how to enable policy change in institutions (13). New resources also continue to emerge (14,15).
3. A report identifying areas where policies are lacking, to catalyse efforts (16).
4. Case studies of how four research organisations have implemented relevant policies (17).

These documents illustrate that there is no single, simple answer as to how to incorporate recognition and support for research software and its personnel into institutional policy. However, policy change is needed in policies related to at least the following categories:

1. Research outcomes/output
2. Copyrights and software licensing
3. Open access and access to research
4. Open science
5. Intellectual property
6. Open source software
7. Research data
8. Research infrastructure
9. Research ethics and integrity
10. Research skills and training
11. FAIR research outputs
12. Privacy and protecting human subjects
13. Research software impacting education
14. Research assessment reform
15. Diversity, equity and inclusion in research

The PRO4RS WG uses the definition of research software from the FAIR for Research Software Principles as including:

> Source code files, algorithms, scripts, computational workflows and executables that were created during the research process or for a research purpose. Software components (e.g., operating systems, libraries, dependencies, packages, scripts, etc.) that are used for research but were not created during or with a clear research intent should be considered software in research and not Research Software. This differentiation may vary between disciplines (18).

Policies that support research software (and/or open source software (OSS)) need to address how to govern the creation, maintenance, acquisition, use, distribution and management of software within their environment, as well as to recognise such work. The answer is not necessarily a new policy on research software, although this can be very beneficial. Another option is to instead include as an element of a number of other policies, or both approaches can be used. The solution is always context dependent. Regardless of the approach, policies need to cover multiple aspects of management. The PRO4RS WG utilised a range of categories for their analysis, including research skills and training, FAIR research outputs,  and research assessment reform (16).

### 4. Broader Initiatives to Support Relevant Policy Development

The PRO4RS WG analysis of policies found that while many institutions include research software in policies on infrastructure, licensing, or open science, few include it substantively in research assessment policies. Yet there are a range of broader initiatives around reform of research assessment that are championing recognition of a range of open science outputs, including software and code. Science Europe's analysis of approaches to research assessment found that the most widely recognised elements of open science as part of the track record assessment of researchers are open access to research articles and books, and FAIR and open research data. However, 62.5% of the 16 organisations responding did include as an element open source research software, code, and tools (19).

This is to expected, as the many hundreds of institutions (and countries) that have signed the San Francisco Declaration on Research Assessment (DORA), and/or are members of the Coalition for Advancing Research Assessment (CoARA), should be in the process of implementing policies that support research outputs including research software. For example, DORA includes as a general recommendation for institutions that research assessment should consider the value and impact of all research outputs, including software. Similarly, the CoARA Commitments include that assessment facilitates the recognition of diverse roles, including software engineers. However, the lack of suitable metrics for measuring research software impact is a broader issue that makes implementation difficult. A particular challenge is the long established approaches used for traditional academic quality and impact measures. Advocacy and support for change are being pursued within an environment shaped by an extensive infrastructure developed over many decades, or even hundreds, of years. This infrastructure is embedded into everyday academic processes, underpinning everything from journals and conferences, to academic hiring and promotion procedures, through to the work of publishers and funders. There is much discussion about whether existing metrics around metrics like citations and journal or conference rankings are correct and fair, and support the development of the best possible research outputs. At the same time, their longstanding nature means they are well understood and very easy to fall back on in the absence of alternatives. At present, some of the more easily accessible metrics to assess the importance and quality of software include things like number of downloads and GitHub stars or equivalents in other source code repositories. However, it has been demonstrated that such metrics are unreliable in this context, with, for example, GitHub stars being used as a way to simply put a placeholder or bookmark on a project, to thank the project's developer(s); as a result of community conventions; or even being purchased to falsely imply importance (20,21). This perhaps helps to highlight the challenge faced by important initiatives such as DORA and CoARA.

Other international policy works offer similar support for inclusion of research software in the reform of research assessment. Science Europe highlights research software as an emerging element of open science policies and practices, and makes recommendations to funders and RPOs on developing and aligning appropriate policies (22). Policy recommendations for US federal government policymakers suggest that policy should take concrete steps to ensure that research software is openly accessible and reusable (23). Ten simple rules for funding scientific OSS argues for focus on issues including scholarly credit, unique forms of labor, and maintenance (24); and the Framework for Managing University OSS recommends creation of pathways for recognition of OSS contributions (25). Finally, the ADORE.software Toolkit, which supports implementation of the Amsterdam Declaration on Funding Research Software Sustainability, includes examples of national and funder policies and relevant resources for research software personnel (26).

**5. Policies, Guidelines and Frameworks to Support Research Software**

This broader landscape analysis indicates that more research institutions should already be actively supporting recognition of research software personnel in recruitment and career advancement than the database of policies seemed to indicate. Consequently, additional resources that contribute to valuing and recognising research software, such as guidelines and frameworks, are also provided in this section. At the research institution level, there are relevant guidelines and frameworks that may provide details on how to implement relevant practices. These may exist even if policies do not, as some research institutions may begin with flexible guidelines rather than formal policies. This may also indicate that policies exist that are not publicly accessible, and thus not discovered in this analysis. And even if an institution does have public policies that support research software and its personnel, these often lack detail on how to operationalise the policy.

Research to identify relevant guidelines, frameworks, and case studies of operationalising recognition for research software outputs in hiring and promoting identified valuable work including the following at the general level:

- The German Psychological Society presents principles of responsible research assessment in hiring and promotion and suggests an assessment procedure that combines the objectivity and efficiency of indicators with a qualitative, discursive assessment of shortlisted candidates (27).
- On the evaluation of research software includes a simplified and flexible protocol concerning research software evaluation: the CDUR procedure (citation, dissemination, use and research) which is intended to provide insight in the relationship between software, its development, its use, and its scientific impact (28).

In terms of specific examples from research institutions, the PRO4RS WG's case studies include the Monash University Business School's Research Software Standards (note that the final policy is not publicly available). This policy proposes that high quality OSS should be treated as a research output by the faculty assessment committee, and provides guidelines to assist staff in making their submission. The Helmholtz Association in Germany is also leading in this area; monitoring of a simple indicator for software was initiated after planning to incorporate software as a new indicator in research evaluation (29). The Helmholtz Research Software Directory is another example of an approach to raising the profile of institutional research software that also supports and demonstrates the need for relevant policies.

At the national and international policy level, broader initiatives that can help support specific policy gaps around recognition and valuing of personnel in institutions include:

- OpenAIRE's model policy recommends that RPOs establish reward mechanisms for researchers using open science practices, with examples including sharing provisional results through open platforms, using open software and other tools (30).
- The OPUS Research Assessment Framework proposes how to assess researchers in an academic context (31), including detailed work on research software indicators (32).
- The RDA SHAring Rewards & Credit (SHARC) Interest Group has developed templates for FAIRness evaluation criteria for researcher self-assessment, which include criteria relating to creation of software management plans (33).

**6. Recommendations**

The PRO4RS WG proposes a common framework for approaching policy support for research software and its personnel, through a range of layers as shown in figure 1. In the central layer, it can be useful to develop a policy specifically on research software as a scholarly output and include the role of its personnel. From this starting point, other policies can be built to align with this (and vice versa). However, the concept of a research software-centric policy is optional, and many institutions omit this layer. The middle layer focuses on ensuring that other policies align with the central layer and/or each other. This may include policies on open science; legal, ethical and secure use; protection of intellectual property rights; interaction with user and developer communities; and how research software engineering skills and outputs are recognised in hiring, evaluation and promotion; etc. Many research institutions may already have some of these policies, but usually lack a complete, coordinated approach. The outer layer encourages alignment with the mechanisms that enable policy that supports research software to be implemented in practice, such as relevant guidelines and frameworks.

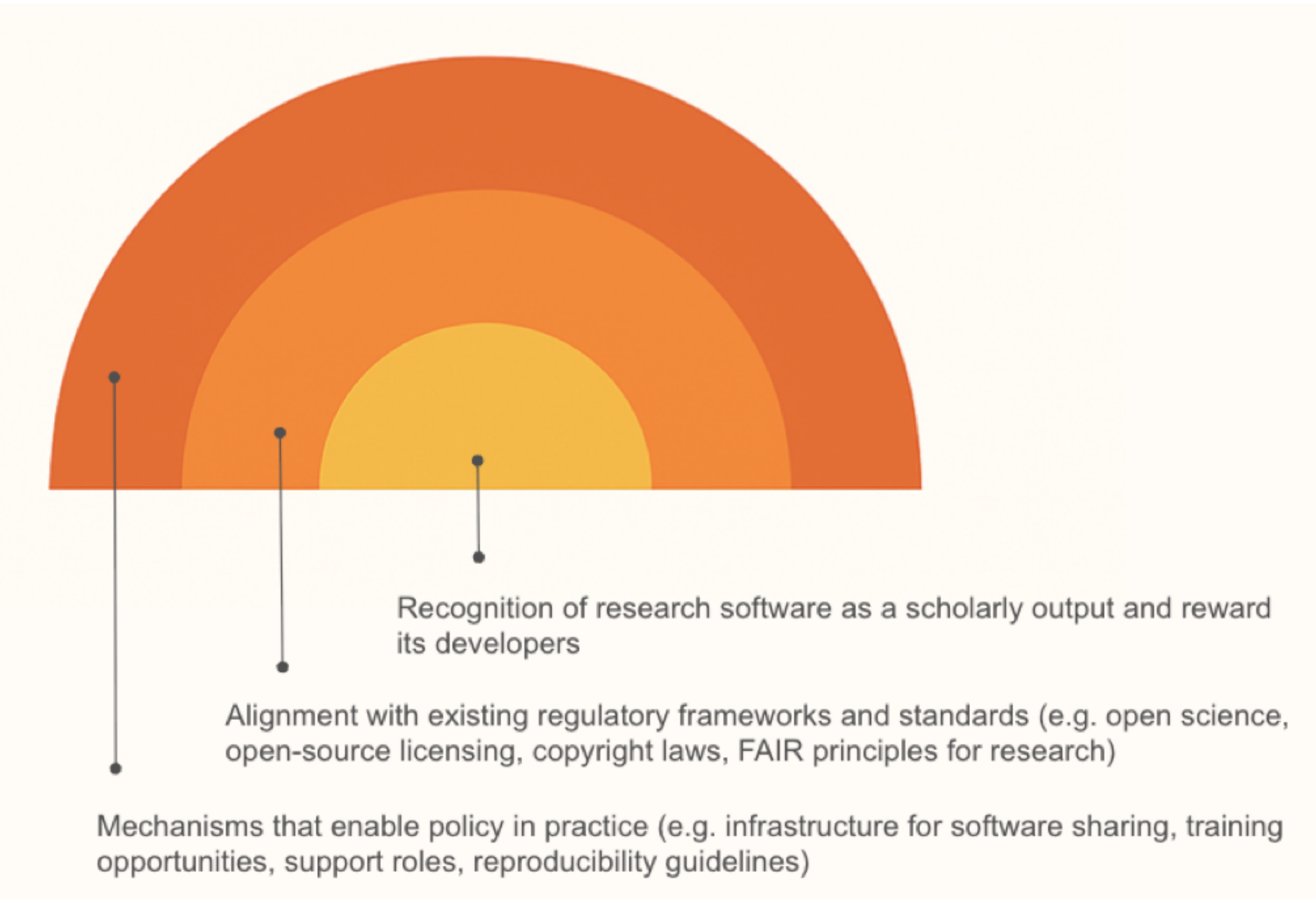

Figure 1: A common framework for approaching policy support for research software and its personnel. Image source: Pedro Hernández Serrano, CC-BY 4.0

It can be very useful for research institutions to start by understanding which of these pieces they already have - overall research software policy in research institutions is a highly fragmented landscape. Some other useful actions may include becoming signatories to the Amsterdam Declaration on Funding Research Software Sustainability and DORA; and members of CoARA (if not already).

Alongside the value that research institutions can gain from developing a research software policy, PRO4RS WG has highlighted a set of actions that both institutional leadership and members of the 'grass roots' community of researchers and technical staff at institutions can undertake to help promote and support the development of policies. These are presented here as a set of three recommendations:

- **Highlight the volume of research software generated within a research institution**, for example, collected using an institutional GitHub organisation to host code, or through deployment of a research software directory where institutional software outputs can be listed.
- Help to **develop recognition of the important role that research software plays** in modern research through advocacy and engagement activities.
- **Engage with key institutional stakeholders**, using the resources for supporting policy change detailed in the PRO4RS WG report, to promote the inclusion of research software policy in a wider suite of research-related policies supporting research quality and excellence.

## 7. Conclusion

Work undertaken by the PRO4RS working group has highlighted the important role that research software policies can play in supporting research. Such policies can help to ensure the use of technical best practices, inform institutional training programmes and underpin the quality of software-based research methods and outputs. At the same time, our work has highlighted that the number of institutions with formal research software policies is still very small.

It is important to remember that research institutions have a key role in changing the research landscape. A key motivation for institutions is often to develop and implement policies to comply with funding agency, government, and disciplinary requirements and norms around research output management. However, research institutions play a critical role as the employers of the personnel who develop and maintain research software; consequently their involvement is critical. And by establishing solid policies for research software management, institutions can promote good research practices, enhance collaboration, foster reproducibility, develop better researchers, and maximise the value to the institution and the impact of efficient and reliable scientific research. By excelling in this area, institutions can enhance their research impact, contribute to raising standards across the wider research community, and raise their profile in comparison with other institutions.

## Acknowledgement

This project has received funding from the European Union’s Horizon Europe research and innovation programme under grant agreement No 101094406.